\documentstyle[a4,11pt]{article}
\input{epsf}
   
 \newcommand{\zd}{\mbox{$\delta$}} 
\newcommand{\ze}{\mbox{$\epsilon$}}

\newcommand{\zs}{\mbox{$\sigma$}}

\newcommand{\xa}{\alpha} \newcommand{\xb}{\beta}

  \newcommand{\xg}{\gamma} 
   
\newcommand{\xl}{\lambda} \newcommand{\xm}{\mu} \newcommand{\xn}{\nu}

\newcommand{\xG}{\Gamma}

\newcommand{\scr}[1]{\mbox{$\cal #1$}}
\newcommand{\itvec}[1]{\mbox{\boldmath ${#1}$}}
\newcommand{\itv}[1]{\mbox{\boldmath ${#1}$}}
\newcommand{\itvx}[1]{\small\mbox{\boldmath ${#1}$}}

\newcommand{\JETP}[3] {Sov. Phys.\ JETP \ {\bf{#1}}, {#2} ({#3})}
\newcommand{\JLTP}[3] {J.\ Low Temp.\ Phys.\ {\bf{#1}}, {#2} ({#3})}
\newcommand{\JPC}[3] {J.\ Phys.\ C {\bf{#1}}, {#2} ({#3})}
\newcommand{\JPF}[3] {J.\ Phys.\ F {\bf{#1}}, {#2} ({#3})}

\newcommand{\JPSJ}[3] {J.\ Phys.\ Soc.\ Jpn.\ {\bf{#1}}, {#2}  ({#3})}
\newcommand{\NPB}[3] {Nucl.\ Phys.\ B {\bf{#1}}, {#2}  ({#3})}

\newcommand{\PR}[3] {Phys.\ Rev.\ {\bf{#1}}, {#2}  ({#3})}

\newcommand{\PRB}[3] {Phys.\ Rev.\ B {\bf{#1}}, {#2}  ({#3})}
\newcommand{\PRL}[3] {Phys.\ Rev.\ Lett.\ {\bf{#1}}, {#2}  ({#3})}

\newcommand{\RMP}[3] {Rev.\ Mod.\ Phys.\ {\bf{#1}}, {#2}  ({#3})}
\newcommand{\SSC}[3] {Solid State Commun.\ {\bf{#1}}, {#2}  ({#3})}
\newcommand{\ZPB}[3] {Z.\ Phys.\ B {\bf{#1}}, {#2}  ({#3})}

\begin{document}

\title{Perturbative Renormalization of
Multi-Channel Kondo-Type Models}
\author{Yoshio Kuramoto\thanks{e-mail:
kuramoto@cmpt01.phys.tohoku.ac.jp}\\
Department of Physics, Tohoku University, Sendai 980-77, Japan}

\maketitle

\begin{abstract}
{
The poor man's scaling is extended to higher order by the use of the open-shell Rayleigh-Schr\"{o}dinger perturbation theory. 
A generalized Kondo-type model with the SU($n$)$\times$SU($m$) symmetry is proposed and renormalized to the third order.
It is shown that the model has both local Fermi-liquid and non-Fermi-liquid fixed points, and that the latter becomes unstable in the special case of $n=m=2$.
Possible relevance of the model to the newly found phase IV in Ce$_{x}$La$_{1-x}$B$_6$ is discussed.
}
\end{abstract}

\section{Introduction}

In the renormalization theory one seeks a small parameter to control perturbative calculation of physical quantities.
In the Kondo model, however, one encounters unmanageable divergence for physical quantities because the fixed point of the model corresponds to the strong coupling limit.  
Thus the numerical renormalization group has proved useful for quantitative study of the Kondo effect \cite{Wilson}.
On the other hand, the multi-channel Kondo model has a nontrivial fixed point at the exchange interaction $J_c$ which becomes small as the number of channels increases.
Then the model has a controlled perturbation expansion from the limit of large degeneracy of conduction bands \cite{Nozieres-Blandin,Cox-Zawadowski}.
It is highly desired to have a simple scheme to perform perturbative renormalization to any desired order.  
This paper proposes an extension of the poor man's scaling to higher orders by the use of the Rayleigh-Schr\"{o}dinger perturbation theory together with the open-shell formalism.

Another motivation of the present study is due to observation of a strange phase (called IV) in Ce$_x$La$_{1-x}$B$_6$ \cite{Nakamura,Sera,Tayama}.
The magnetic susceptibility shows a cusp on entering the phase IV from the paramagnetic phase with decreasing temperature.  
This suggests that  the N\'{e}el state is present here.  In contrast to the phase III which has both antiferromagnetic and quadrupole orders, however, the phase IV has a very small magnetic anisotropy and almost no magneto resistance.
We recognize the importance of the orbital Kondo effect which is active even in the presence of a spin ordering.
In order to inspect the importance one needs to know how the intersite interaction and the on-site orbital and spin Kondo effects compete. 
As the first step toward this direction, 
we study the simplest impurity model that displays both the orbital and spin Kondo effects with special attention to the combined spin and orbital symmetry of the 4$f$ shell.

In conventional perturbation theory for Kondo-type models one introduces fictitious fermions or bosons to represent the localized spin \cite{Abrikosov,Barnes,Coleman}.
Then one computes some lower-order Feynman diagrams in order to apply the multiplicative renormalization \cite{Abrikosov-Migdal,Fowler-Zawadowski}.
However, it is known that there is no linked-cluster expansion in this pseudo-particle theory \cite{Keiter}.  
Thus in higher-order it is difficult to accomplish proper counting of perturbation processes.
To overcome the difficulty several methods have been proposed.  Among them is the resolvent formalism or the Brillouin-Wigner-type perturbation theory, which does not rely on the linked-cluster property \cite{Keiter-Kimball,Kuramoto,Grewe,Bickers}.
Another elegant way is to introduce a particular value of the imaginary chemical potential for the fictitious fermions.  Then unphysical contributions cancel each other \cite{Popov,Gan}.  We note that the latter method  works only for the magnetic impurity with spin 1/2.

In doing perturbative renormalization of the Kondo model, the ``poor man's scaling" of Anderson \cite{Anderson} provides the simplest framework for practical computation.
If one wants to proceed to higher-order renormalization, however, extention of the scaling method is not straightforward \cite{Solyom-Zawa,Hewson}.  
In this paper, we propose an alternative approach to extend the poor man's scaling to arbitrary high orders. 
In contrast to the original idea of using the invariance property of the $t$-matrix, the present theory considers the effective Hamiltonian which gives the same energy spectrum as the original one at each order of renormalization.
In deriving the effective Hamiltonian we use the Rayleigh-Schr\"{o}dinger perturbation theory in the open-shell formalism, which is popular in quantum chemistry and nuclear physics \cite{Lindgren,Brandow}. 
We demonstrate the usefulness of our 
scheme by computing the third-order term in the renormalization group.

The organization of the paper is as follows:  In the next section we review the effective Hamiltonian formalism by the use of the Rayleigh-Schr\"{o}dinger perturbation theory.  
Section 3 performs renormalization to third order explicitly for the multi-channel Kondo model.  
We explain the notion of the folded diagram which is very convenient to exploit the linked cluster property.
These two sections do not give new results to the Kondo problem itself, but explains the formalism and demonstrates its simplicity.
In Section 4 we introduce an SU($n$)$\times$SU($m$) model where $n$ denotes the number of spin degrees of freedom, and $m$ does the orbital one.  This model is different from the SU($n$)$\times$SU($m$) Anderson model already discussed in the literature \cite{Cox,Kroha}.
Our model includes both the Fermi-liquid and non-Fermi-liquid fixed points, and  reduces to the multi-channel Kondo model in a special case.
The renormalization to the third order is performed for general $n$ and $m$ by the use of the Lie algebra.
We give detailed analysis of each fixed point and the renormalization flow.
The final section summarizes the paper and describes possible further development.

\section{Effective Hamiltonian in the Model Space}
\subsection{Projection to the model space}

As a preliminary to perturbative renormalization we introduce the projection operator $P$ to the model space $M$.  
We divide the Hamiltonian $H = H_0+V$ such that the unperturbed part $H_0$ commutes with $P$.  
In many cases we are interested only in certain lower part of the full spectrum of $H$.   
The effective Hamiltonian $H_{eff}$ is designed to give the identical spectrum in the model space as that of $H$.   Namely if $\psi_i$ is an eigenfunction of $H$ with the eigenvalue $E_i$, we require
\begin{equation}
H_{eff}P\psi_i = E_i P\psi_i.
\end{equation}
To construct $H_{eff}$ it is necessary to introduce the wave operator $\Omega$ which reproduces $\psi_i$ from  $P\psi_i$, namely 
$\psi_i=\Omega P\psi_i$.
Note that the operator product $\Omega P$ is not an identity operator since $P$ does not have an inverse. We emphasize that the above reproduction is possible only because $\psi_i$ is an eigenfunction of $H$.   

In the Brillouin-Wigner perturbation theory the wave operator can be derived in the simple form \cite{Lindgren}:
\begin{equation}
\Omega (E_i) = \sum_{n=0}^\infty \left(\frac{1}{E_i-H_0} QV\right)^n, 
\end{equation}
where $Q=1-P$ is the projection operator to the space complementary to the model space.
In spite of its simple appearance, 
the energy dependence of the wave operator makes it rather 
inconvenient for term-by-term perturbation theory in higher orders. 
On the other hand, the Rayleigh-Schr\"{o}dinger perturbation theory is able to derive $\Omega$ and hence $H_{eff}$ in a form which involves only eigenenergies of $H_0$ and is free from the unknown energy $E_i$.
We write the Schr\"{o}dinger equation in the two forms:
\begin{eqnarray} 
(E-H_0) \psi & = & E\psi -H_0\Omega  P\psi = V\Omega P\psi,
\label{eq:H-Omega} \\
\Omega (E-H_0)P\psi & = & E\psi -\Omega H_0 P\psi = \Omega PV\Omega P\psi, 
\label{eq:Omega-H} 
\end{eqnarray}
where the index $i$ to specify an eigenstate is omitted.
The latter equation is obtained by application of $P$ on both sides of eq.(\ref{eq:H-Omega}) and further application of $\Omega$.
Subtracting eq.(\ref{eq:Omega-H}) from eq.(\ref{eq:H-Omega}), 
we obtain
\begin{equation}
[\Omega, H_0] P\psi =  (1-\Omega P)V\Omega P\psi .
\label{eq:RS-com}
\end{equation} 
We make the power series expansion:
$$ \Omega = \Omega_0 +\Omega_1 +\Omega_2 +\ldots,$$ where $\Omega_n$ denotes  the $O(V^n)$ contribution with $\Omega_0=1$.
Comparing  the terms with the same order of magnitudes on both sides of eq.(\ref{eq:RS-com}), we obtain
\begin{equation}
[\Omega_n, H_0]  =  QV\Omega_{n-1} -  \sum_{j=1}^{n-1}\Omega_{j}PV\Omega_{n-j-1}.
\label{eq:Omega_n}
\end{equation}
The effective Hamiltonian is accordingly expanded as
\begin{equation}
H_{eff} = P(H_0 +V)P+H_2+H_3+\ldots,
\end{equation}
with $H_n = PV\Omega_{n-1}P $ for $n\ge 2$.   

Matrix elements of some lower-order terms are explicitly given by
\begin{eqnarray}
\langle a|H_2| b\rangle &=& \langle a|V(\ze _b-H_0)^{-1}QV| b\rangle,\\ 
\langle a|H_3| b\rangle &=& \langle a|V\frac{1}{\ze _b-H_0} QV\frac{1}{\ze _b-H_0} QV| b\rangle - 
\sum_c \langle a|V\frac{1}{\ze _b-H_0} \frac{1}{\ze _c-H_0} QV |c\rangle 
\langle c|V| b\rangle,
\label{eq:RS-third}
\end{eqnarray}
where states such as $|a\rangle, |b\rangle$ and $|c\rangle$ belong to the model space with $H_0|b\rangle = \ze _b|b\rangle$ etc.
We note that the third-order Hamiltonian is not Hermite.
It is possible to convert it to a Hermitian operator by a similarity transformation \cite{Brandow}.

\subsection{Linked cluster property}

In the Rayleigh-Schr\"{o}dinger perturbation theory, two kinds of terms appear in $H_n$ with $n\ge 3$; the first kind takes the simple form 
$[(\ze _b-H_0)^{-1}QV]^n$ 
and the other kind of terms have less simple form and originate from the second term in eq.(\ref{eq:Omega_n}).
The latter terms in fact play an important role in leading to the linked-cluster expansion.
Namely, among all processes appearing as a result of Wick decomposition,
unlinked diagrams are cancelled by the second kinds of terms \cite{Lindgren}.  
Here the linked or unlinked diagrams have the same meanings as those in the Feynman diagram method.  

The Rayleigh-Schr\"{o}dinger theory has two cases to treat: the simpler one is the closed shell case where a large energy gap separates the occupied and unoccupied fermion orbitals. 
The other one is the open shell case where in addition to the stable core orbitals, a part of nearly degenerate orbitals are occupied.  
The latter orbitals are called valence orbitals, and constitute the model space.
The essential point of applying the open-shell formalism to Kondo-type models is to regard the local electron as belonging to the set of valence orbitals. 
In order to utilize the linked-cluster property of the perturbation theory, we 
remove the restriction that the number of local electrons be unity at any stage of intermediate states.
The final form of the effective Hamiltonian conserves the number of local electrons in each order of perturbation.  
Hence in the model space with one and only one local electron which obeys either the Fermi or Bose statistics, one can reproduce the actual situation in the Kondo-like impurity model with the local number constraint.
Note that this reproduction is possible because we are working with the canonical ensemble instead of the grand canonical one. 

In order to deal with the energy denominators of terms such as the second term in eq.(\ref{eq:RS-third}), the concept of the ``folded diagram" is useful \cite{Lindgren,Brandow}. 
Namely each time the projection operator $P$ appears in the expansion, 
one draws a diagram where the direction of propagation is reversed.
Then the energy denominator is associated with difference of left-going energies and the right-going ones just as in the first term of eq.(\ref{eq:RS-third}). 
By comparing the energy denominators given by the second term of eq.(\ref{eq:RS-third}), and those given by the rule for the folded diagrams,  one can confirm that both give identical results.

\section{Renormalization of the Multi-Channel Kondo Model}

As the simplest application of the present method,
we consider the multi-channel Kondo model defined by
\begin{equation}
H_{K} =  \sum_{{\itvx k}\sigma}\sum_{l=1}^n \ze_{{\itvx k}} c_{{\itvx k} l\sigma}^\dagger c_{{\itvx k} l\sigma} + J\itv S\cdot\itv s
\label{eq:Kondo}
\end{equation}
with channel index $l$.
Here $\itv S$ is the localized spin with magnitude 1/2 and 
\begin{equation}
\itvec s = {1\over {2N}}\sum_{\itvec {k k'}}\sum_{\xa\xb}\sum_{l=1}^n c_{\itvec k l\xa}^\dagger \itvec\zs _{\xa\xb} c_{\itvec k' l\xb},
\end{equation}
with $N$ being the number of $\itvec k$-states and 
$\itv\sigma$  the Pauli matrix.
It is known that the fixed point of the model gives the non-Fermi liquid for $n \geq 2$ \cite{Nozieres-Blandin}.
In order to perform renormalization explicitly we assume the constant density of states for $n$ conduction bands each of which per spin is given by
$$
\rho_c (\ze )= (2D)^{-1}
$$ for $|\ze | <D$ and zero otherwise.
The model space consists of the local spin and such conduction electrons whose energy $\ze$ is within the range $[-D+|dD|,D-|dD|]$ where $dD (<0)$ is the infinitesimal change of the cut-off energy.
In order to simplify the diagrams we define the vacuum as the Fermi sea of non-interacting conduction electrons, and work with the particle-hole picture for excitations.

\begin{figure}
\begin{center}
\epsfxsize=12cm \epsfbox{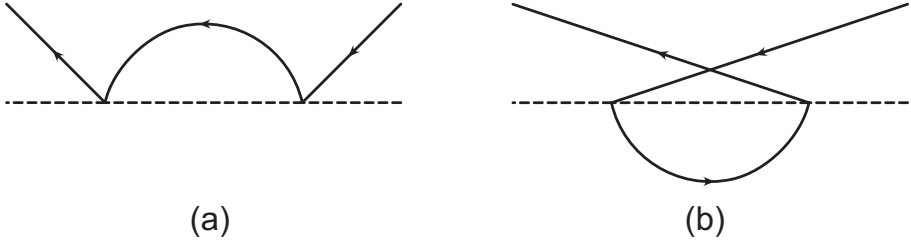}
\end{center}
\caption{Scattering processes in second order.  The solid line shows a conduction-electron state, while the dashed line the local electron.
 The projection operator $Q$ requires the intermediate conduction-electron states to have energies near the band edges.
}
\label{fig:2nd}
\end{figure}

In the lowest order,  we consider the two processes shown in Fig.\ref{fig:2nd}.  
We associate the energy $\pm D$ to the intermediate conduction states which belong to the orthogonal subspace projected by $Q$.
As is clear from the direct calculation,  
the effective exchange interaction depends on the energy of conduction electrons in contrast to the bare interaction.  
We replace this energy dependent interaction by a constant, and continue the renormalization with the same form of the effective Hamiltonian.
Namely it is assumed that  the dimensionless effective interaction $g(D)$ with a given cut-off energy $D$ is related to another with a different cut-off energy $D'$ by a multiplicative factor $z(g, D'/D)$.  
We require
\begin{equation}
 g(D') = z(g, D'/D) g(D).
\end{equation}
This multiplicative property can be expressed equivalently in the differential form:
\begin{equation}
 \frac{\partial g}{\partial l} =\beta(g),
\end{equation}
where $l=\ln D$ and the beta-function in  the right-hand side does not depend explicitly on $l$.
This scheme is called the logarithmic approximation in the literature \cite{Nozieres69}.
In the effective Hamiltonian approach,  
the logarithmic approximation requires that the infinitesimal change of the effective interaction through one step of renormalization is proportional to $D^{-1}$ and is independent of the external energy.  
In the lowest order the independence is approximately satisfied for
the external energy much smaller than $D$.  To proceed similarly in the third order, we set $\ze _b =0$ in eq.(\ref{eq:RS-third}) and deal with
the effective Hamiltonian:
\begin{equation}
 H_3^L = V(H_0^{-1}QV)^2 +\sum_c VH_0^{-1}(\ze _c-H_0)^{-1}QV|c\rangle\langle c|V.
\label{eq:H3log}
\end{equation}
 We refer to the literature \cite{Nozieres69} for detailed discussion of  the logarithmic approximation by a different formalism.

Before performing the third-order renormalization for the Kondo model,
it is instructive to consider first the potential scattering from the impurity instead of the exchange scattering.  If the particle-hole symmetry is present in the conduction band, we can generate a particle-hole conjugate diagram from any given diagram by reversing the direction of all the conduction-electron lines.  
Then the incoming conduction-electron line which corresponds to the annihilation operator in the effective Hamiltonian and the outgoing one for the creation operator are interchanged.  
The diagrams  shown in Fig.\ref{fig:2nd} represent the 
simplest example of a pair of conjugate diagrams.
For the potential scattering the two diagrams in Fig.\ref{fig:2nd} combine to zero.   
This is because an extra interchange of the incoming and outgoing fermion lines are involved in constructing the conjugate diagrams.
To the contrary, the non-commutativity of spin operators lead to nonzero result from the sum of two diagrams in Fig.\ref{fig:2nd}.  

\begin{figure}
\begin{center}
\epsfxsize=6cm \epsfbox{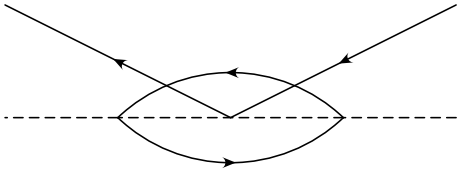}
\end{center}
\caption{
Exchange scattering processes in the third order.  
The projection operator $Q$ requires one of two conduction-electron states in the loop to have energies near the band edges.
}\label{fig:3rdvtx}
\end{figure}

\begin{figure}
\begin{center}
\epsfxsize=6cm \epsfbox{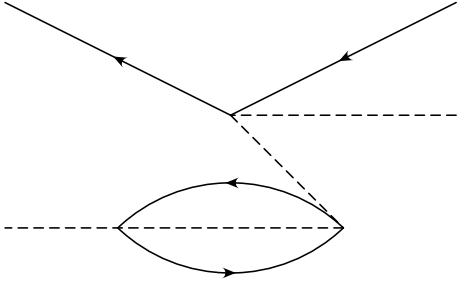}
\end{center}
\caption{The third-order folded diagram. The assignment of energy denominators are explained in the text.
}\label{fig:folded}
\end{figure}

In the third-order renormalization we first consider such contributions that become dominant for large $n$.
These are diagrams which have a loop of conduction electrons as shown in Figs.\ref{fig:3rdvtx} and \ref{fig:folded}, since 
each loop of conduction electron lines acquires the factor $n$ by summing over degenerate orbitals.  
The diagram shown in Fig.\ref{fig:3rdvtx}  corresponds to the first term in eq.(\ref{eq:H3log}).
The folded diagram shown in Fig.\ref{fig:folded} corresponds to the wave-function renormalization in the Green function formalism.
With the notion of folded diagrams one can confirm that the cancellation of potential scattering diagrams persists at least to the third order.
Namely the diagram shown in Fig.\ref{fig:folded} has the same magnitude but has the sign opposite to the one shown in Fig.\ref{fig:3rdvtx}. 
Thus the two contributions cancel each other for the case of potential scattering.
In both Figs.\ref{fig:3rdvtx} and \ref{fig:folded}, the energy denominators are obtained by associating the excitation energy $D$ with one of two lines in the electron loop.  
The contribution is given by
\begin{equation}
\int_{-D}^0 \frac{d\ze '}{(-D+\ze ')^2}+
\int_0^D \frac{d\ze}{(-D-\ze )^2} = \frac 1D.
\label{eq:energy denominator}
\end{equation}
The product of the spin operators takes the form
$
S^\xa S^\xb S^\xg {\rm Tr}(s^\xa s^\xg) s^\xb
$
where the trace is over the spin states of conduction-electrons.
On the other hand the folded diagram has the same energy denominators as given by eq.(\ref{eq:energy denominator}), and
the spin part is given by 
$
S^\xa S^\xg S^\xb {\rm Tr}(s^\xa s^\xg) s^\xb .
$
Then contributions from the two diagrams combine to give
\begin{equation}
\sum_{\xa\xb\xg} S^\xa [S^\xb, S^\xg] {\rm Tr}(s^\xa s^\xg) s^\xb = -\frac 12 \itv{S\cdot s}, 
\end{equation}
where we use the identity
$
[S^\xa, S^\xb] =  i \ze _{\xa\xb\xg}S^\xg 
$
with $\ze _{\xa\xb\xg}$ being the completely antisymmetric unit tensor.

As will be shown shortly, all non-loop diagrams in the third order can be neglected for multiplicative renormalization.
Thus we recover the known result  \cite{Nozieres-Blandin}
\begin{equation}
\frac{dg}{dl} = -g^2 + \frac n2 g^3,
\end{equation}
where $g=J\rho_c$.
A trivial fixed point of the model is $g_c=0$ which, however, is unstable.
Another fixed point of the renormalization group is given by $g_c=2/n$ which is small in the case of large $n$.  Thus the fixed point is within the reach of the perturbative renormalization.
Linearization of the scaling equation around $g_c=2/n$ shows that the latter fixed point is stable.

\begin{figure}
\begin{center}
\epsfxsize=6cm \epsfbox{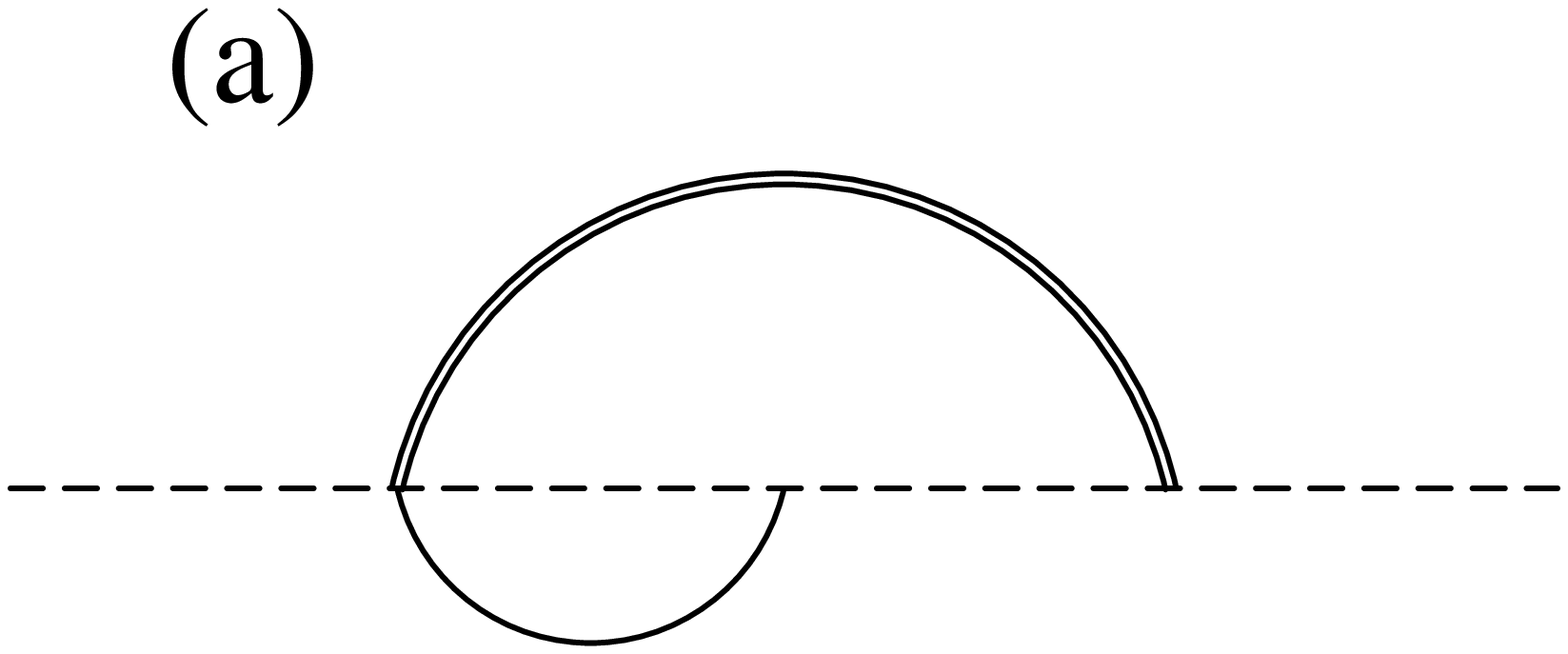}
\epsfxsize=6cm \epsfbox{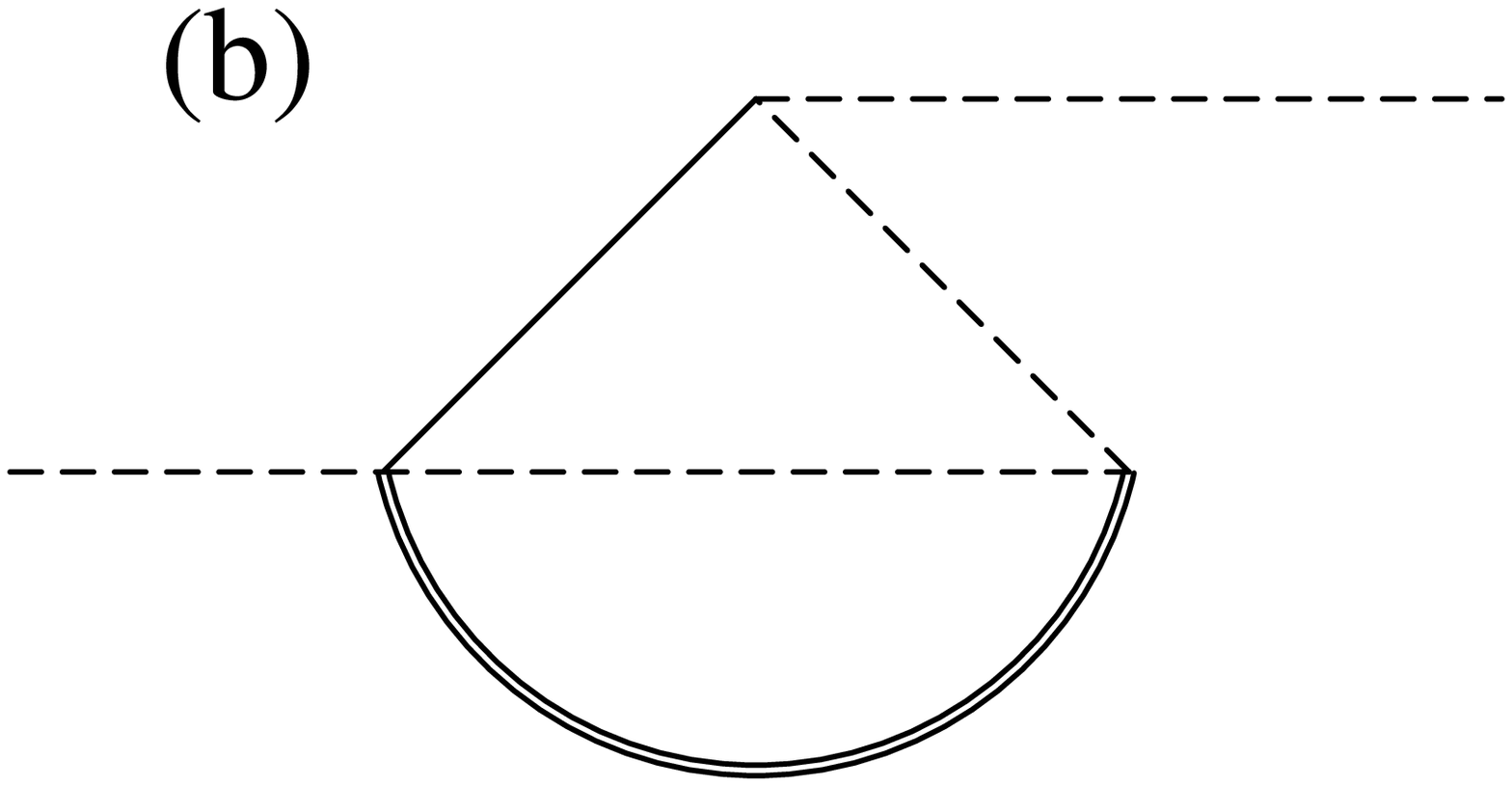}
\epsfxsize=6cm \epsfbox{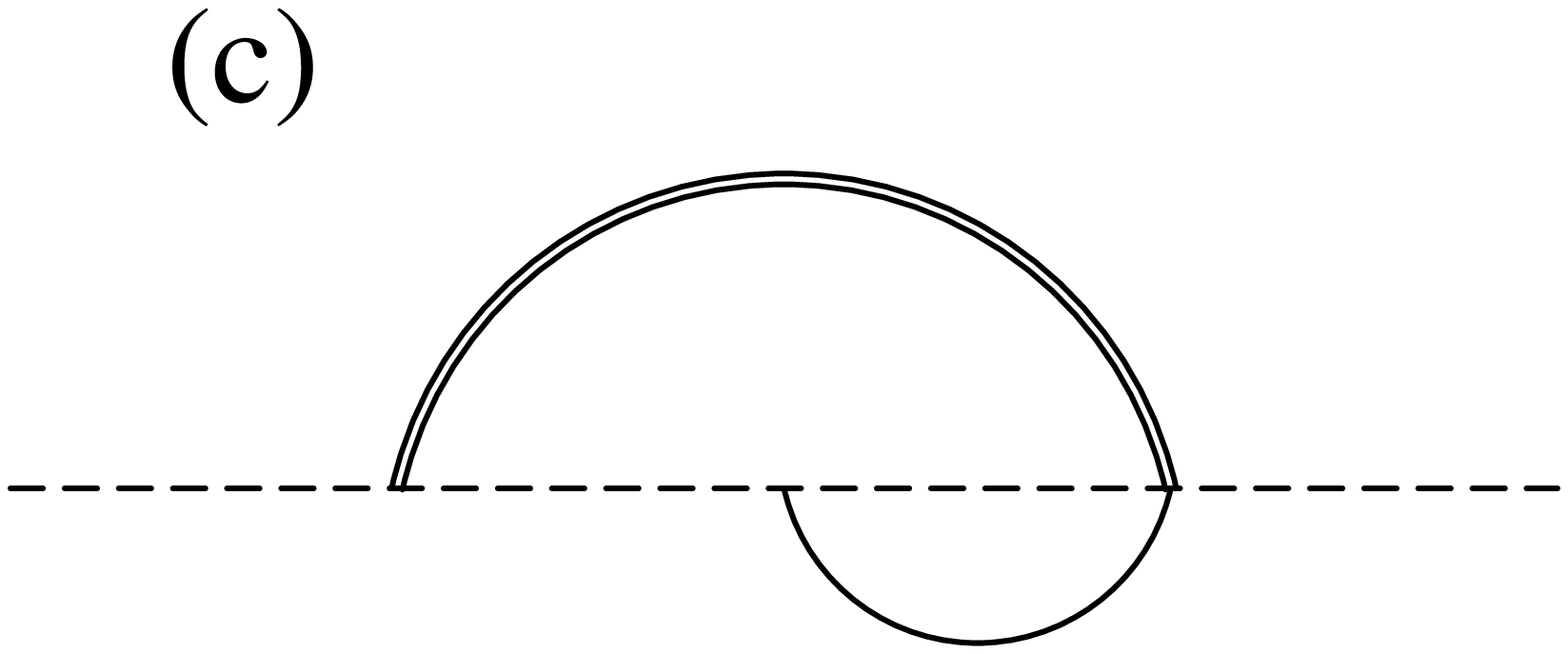}
\epsfxsize=6cm \epsfbox{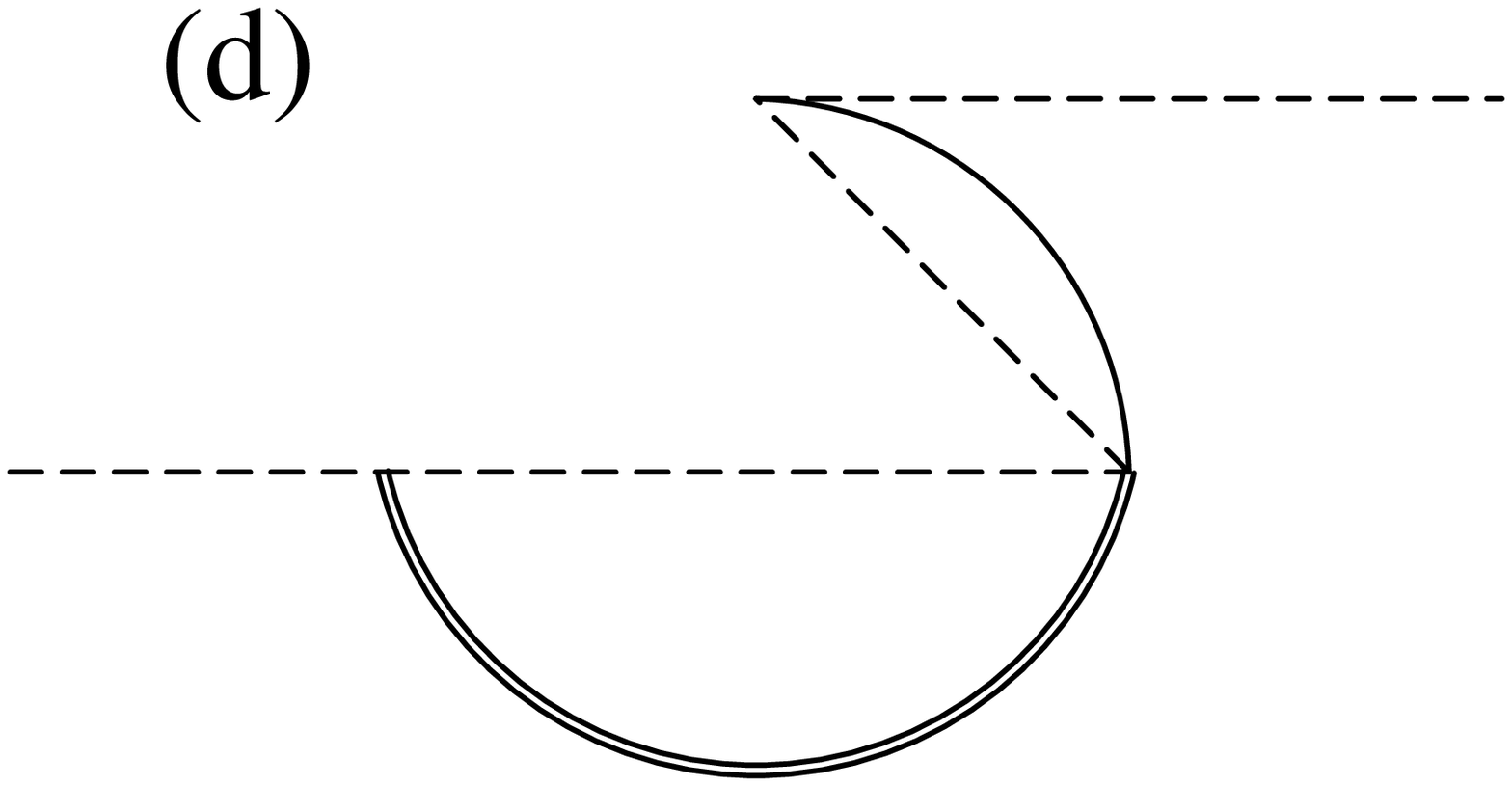}
\end{center}
\caption{
Exchange scattering processes in the third order without a conduction-electron loop.
The projection operator $Q$ requires the double lines to have energies near the band edges. 
There are two ways of associating the direction of arrows for each diagram.
}
\label{fig:3rd}
\end{figure}

Now we turn to non-loop diagrams in the third order.
Figure \ref{fig:3rd} shows such contributions with incoming and outgoing conduction electron lines deleted.  These diagrams belong to the group called the ``parquet'' in the Green function theory \cite{Nozieres69}.
The two ways of attaching directions of arrows correspond to particle-hole conjugate diagrams.  
It can be seen by direct calculation that the diagrams shown in Fig.\ref{fig:3rd}(a) and (b) cancel each other not only for the potential scattering but for the exchange scattering.
This is because both have the same energy denominators and matrix elements but have different signs.
On the other hand, the diagram shown in Fig.\ref{fig:3rd}(c) and (d) do not cancel among themselves.  Moreover the diagram (d) is logarithmically divergent.  This divergence is, however, not related to the Kondo effect since the potential scattering also gives this divergence.  

Let us ask the nature of these contributions to the effective interaction in the third order. 
They have a common feature of being constructed from the second-order diagrams through replacement of a bare interaction vertex by the second-order effective interaction.  
A diagram with such property is called ``reducible''.
We note that the change of the cut-off influences the shift of the ground-state energy, since the band-edge parts are no longer available to the effective Hamiltonian in the next step.
To reproduce the identical energy shift as that of the original model, the effective Hamiltonian should contain a part which compensates the missing energy shift associated with the cut-off region.
In the third order, this compensation comes from the expectation value of the reducible part of the effective interaction. 
In fact a part of the third-order energy shift comes from the diagram 
which is obtained from  Fig.\ref{fig:3rd} (d) by joining the outgoing conduction-electron line with the incoming one.
The resultant energy shift corresponds to the expectation value of the logarithmically divergent effective interaction.
In a similar manner expectation value of the diagram in Fig.\ref{fig:3rd} (c) also compensates another missing part of  the energy shift associated with the cut-off region.

Thus we conclude that in the third-order non-loop diagrams do not contribute to the multiplicative renormalization.
The effective interaction relevant to  the renormalization group comes only from ``irreducible diagrams'' which cannot be constructed from lower-order diagrams through replacement of a bare interaction by a ``vertex part'' in the terminology of the Green function formalism.
In the conventional renormalization theory, the number of diagrams increases considerably as one goes higher order in $g$.
These terms have been discussed in the literature \cite{Abrikosov-Migdal,Gan} with mutually inconsistent results.
The inconsistency may partly be related to non-universal character involved in  higher-order renormalization.
We plan detailed account of $O(g^4)$ renormalization and comparison with available exact results in a future publication.
In this section we reproduced the old result just for the purpose of demonstrating the simplicity of the new formalism.

\section{Renormalization of Generalized Kondo-Type Models}

\subsection{SU(2)$\times$SU(2) Kondo model}
\label{subsec:2 by 2 model}

The orbital dynamics in rare-earth systems such as Ce$_{x}$La$_{1-x}$B$_6$ has remarkable entanglement of magnetic and electric multipoles.
The four-fold degenerate crystal field level $\xG _8$ is often treated in the framework of the SU(4) symmetry.  
However, if one starts with the Anderson model and renormalizes off the charge fluctuation, the multiplet structure of $4f^2$ configurations act differently to the orbital and spin degrees of freedom \cite{Cox-Zawadowski}.  Hence we consider the following model \cite{Pang}.
\begin{equation}
H_{2\times 2} =  \sum_{{\itvx k} l\sigma}\ze_{{\itvx k}} c_{{\itvx k} l\sigma}^\dagger c_{{\itvx k} l\sigma} + H_{ex}
\label{eq:SU(2)Kondo}
\end{equation}
where $l = 1,2$ denotes orbital index.  The exchange part $H_{ex}$ is given by
\begin{equation} 
H_{ex}= J f^\dagger \itv S f\cdot c^\dagger \itv s c+K f^\dagger \itv T f\cdot c^\dagger\itv \tau c+ 4I \sum_{\xm\xn} (f^\dagger S^\mu T^\nu f)( c^\dagger s^\mu \tau^\nu c).
\label{eq:2by2 model}
\end{equation}
where $\itv S$ and $\itv s$ refer to spin matrices and $\itv T$ and $\itv\tau$ do the orbital ones.
They are given explicitly by
\begin{eqnarray}
f^\dagger \itv S f &=&\frac 12 \sum_{l\xa\xb}f^\dagger_{l\xa}\itv\zs _{\xa\xb}f_{l\xb}, \ \ \ 
f^\dagger \itv T f =\frac 12 \sum_{lr\xa}f^\dagger_{l\xa}\itv\zs _{lr}f_{r\xa},\\
f^\dagger S^\mu T^\nu f &=& \frac 14 \sum_{lr}\sum_{\xa\xb}f^\dagger_{l\xa}\zs _{\xa\xb}^\mu \zs _{lr}^\nu f_{r\xb}
\end{eqnarray}
for localized electrons.  Similar notation has also been used for conduction electrons.
There is a constraint to suppress the charge fluctuation of $f$ electrons:
\begin{equation}
\sum_{l\xa}f_{l\xa}^\dagger f_{l\xa} = 1.
\label{eq:constraint}
\end{equation}

In the special case of $J=K=I$, the present model reduces to the Coqblin-Schrieffer model.  To see this we note that the spin permutation operator $\scr P_s$ and the orbital one $\scr P_l$ can be written as
\begin{eqnarray}
\scr P_s = 2f^\dagger \itv S f \cdot c^\dagger \itv s c +\frac 12,
\label{eq:spin permutation}\\
\scr P_l = 2f^\dagger \itv T f \cdot c^\dagger \itv\tau  c +\frac 12 .
\label{eq:orbital permutation}
\end{eqnarray}
Then the exchange part of the Coqblin-Schrieffer model is represented by
\begin{equation}
I f^\dagger  c^\dagger (2\itv S  \cdot\itv s +\frac 12)
(2\itv T \cdot \itv\tau +\frac 12) c f,
\end{equation}
in an abbreviated notation.  This interaction reduces to eq.(\ref{eq:2by2 model}) with $J=K=I$.

Renormalization of the model proceeds in a similar manner as that in the Kondo model.  In order to simplify the notation we use the convention that $J,K$ and $I$ denote the effective interaction and the unit of energy is such that $\rho_c =1$.
In the second order cross terms of $J$ and $K$ vanish because the interactions with them commute each other.  This is not the case with the $I$ term.
In the third order the relevant diagrams are again only those shown in Figs.\ref{fig:3rdvtx} and \ref{fig:folded}.  The vertex parts there can be either $J,K$ or $I$.  On the other hand the energy denominators are the same as in the case of the Kondo model. 
The resultant scaling equations are given by
\begin{eqnarray} 
\frac{\partial J}{\partial l} &=& -(1-J)(J^2+3I^2),\\
\frac{\partial K}{\partial l} &=& -(1-K)(K^2+3I^2),\\
\frac{\partial I}{\partial l} &=& -2I(J+K)+I(K^2+J^2+2I^2).
\label{eq:I-scaling}
\end{eqnarray} 

Let us discuss implication of the scaling equations. 
The set of equations has six fixed points (i) -(iv) as follows:  
\begin{eqnarray*}
\begin{array}{lll}
{\rm (i)}\  J_c=K_c=I_c=0, &  {\rm (ii)}\ J_c=I_c=0, K_c=1, &  {\rm (iii)}\ J_c=1, K_c= I_c=0, \\
{\rm (iv)}\  J_c=K_c=1, I_c=0, & {\rm (v)}\ J_c=K_c=1, I_c=1, &  {\rm (vi)}\ J_c=K_c=1, I_c=-1.
\end{array}
\end{eqnarray*}
The fixed points (ii), (iii) and (iv) correspond to the non-trivial fixed point known for the multi-channel Kondo model \cite{Nozieres-Blandin} which are in fact unstable in the presence of $I$.   
We discuss in more detail the stability of these fixed points in the next section taking a generalized model. 
It is known that the Coqblin-Schrieffer limit of the model does not have the non-Fermi liquid.
Then the fixed point (v) with $I_c=1$ should be an artifact of the third-order scaling.
The correct fixed point is at $I_c=J_c=K_c=\infty$ and gives the local Fermi liquid. 

A remarkable property of the scaling given by eq.(\ref{eq:I-scaling}) is the absence of the $I^2$ term.  
The SU(2) symmetry plays a special role in this absence as will become clear in the next section.
Since the absence means that the sign of $I$ is irrelevant in the present system, 
it suggests that the fixed point (vi) with $I=-1$  also flows to infinity in the exact renormalization group.  
This flow has been confirmed by the numerical renormalization group method \cite{Kusunose-Kuramoto}.  
Namely the spectrum of the model with positive $I$ is the same as that with negative $I$ for the same absolute value.
The flow to the Fermi liquid is also concluded by Pang \cite{Pang}, who started the renormalization from $I=0$. 


\subsection{SU($n$)$\times$SU($m$) model}

Motivated by the previous success of the large $n$ theory for the multi-channel Kondo model \cite{Nozieres-Blandin,Cox-Zawadowski}, we now generalize the model of eq.(\ref{eq:SU(2)Kondo})  to the SU($n$)$\times$SU($m$) symmetry with arbitrary $n$ and $m$.
Before presenting the SU($n$)$\times$SU($m$) model and its renormalization,  
we quote necessary formula for the Lie algebra of the unitary group SU$(n)$.  Let  $X^\xa \ (\xl = 1,2,\ldots n^2-1)$ represent the set of generators of the SU($n$) Lie algebra.  The commutation rule is given by
\begin{equation}
[X^\xa, X^\xb ] = i f_{\xa\xb\xg}X^\xg, 
\end{equation}
where $f_{\xa\xb\xg}$ is called the structure constant,  and is completely antisymmetric against interchange of  a pair of indices.  In the SU(2) case $f_{\xa\xb\xg}$ reduces to the unit tensor $\epsilon_{\xa\xb\xg}$ of the third rank.
Then we require the orthonormality
\begin{equation}
{\rm Tr} (X^\xa X^\xb ) = \frac 12\zd _{\xa\xb},
\end{equation}
which is the generalization of the relation obeyed by the SU(2) spin matrices.
The completeness relation is expressed as
\begin{equation}
\sum_\xa (X^\xa)_{ab}( X^\xa )_{cd} = \frac 12 (\zd _{ad}\zd _{bc} -\frac 1n  \zd _{ab}\zd _{cd} ).
\label{eq:complete}
\end{equation}
In the special case of SU(2), the above relation is translated to that of spin permutation $\scr P_s$ of local and conduction electrons as given by eq.(\ref{eq:spin permutation}).

By analogy with eq.(\ref{eq:2by2 model}) we now introduce the model with
 $n$-fold degenerate spin degrees of freedom, and $m$-fold degenerate orbital degrees of freedom.  The exchange part of the Hamiltonian is given by
\begin{equation} 
H_{ex}= J \sum_\mu (f^\dagger S^\mu f)(c^\dagger s^\mu c)+K \sum_\nu (f^\dagger T^\nu f)(c^\dagger \tau^\nu c)+ 4I \sum_{\xm\xn} (f^\dagger S^\mu T^\nu f)( c^\dagger s^\mu \tau^\nu c).
\label{eq:nbym model}
\end{equation}
where 
\begin{eqnarray}
f^\dagger S^\mu f &=&\frac 12 \sum_{l\xa\xb}f^\dagger_{l\xa} (X^\mu    )_{\xa\xb}f_{l\xb}, \ \ \
f^\dagger T^\nu f =\frac 12 \sum_{lr\xa}f^\dagger_{l\xa} (X^\nu)_{lr} f_{r\xa},\\
f^\dagger S^\mu T^\nu f &=& \frac 14 \sum_{lr}\sum_{\xa\xb}f^\dagger_{l\xa}(X^\mu) _{\xa\xb} (X^\nu) _{lr}f_{r\xb}
\end{eqnarray}
for localized electrons.  There is the constraint given by eq.(\ref{eq:constraint}) as in the case of the SU(2) $\times$ SU(2) system.

In the special case of $I=mJ/2=nK/2$, the present model reduces to the SU($n\times m$) Coqblin-Schrieffer model.  This becomes apparent when the product $\scr P_s \scr P_l$ is written in terms of generators of the Lie algebra.
Namely we can write $\scr P_s$ by using eq.(\ref{eq:complete}) as 
\begin{equation}
\scr P_s =  2\sum_\mu (f^\dagger S^\mu f)(c^\dagger s^\mu c) +\frac 1n,
\end{equation}
and similar generalization of eq.(\ref{eq:orbital permutation}).

The scaling equation up to the third order is given by
\begin{eqnarray} 
\frac{\partial J}{\partial l} &=& -\frac n2 \left(1-\frac m2 J\right)
\left[J^2+4\left( 1-\frac {1}{m^2} \right)I^2\right], 
\label{eq:J-3rd}\\
\frac{\partial K}{\partial l} &=&   -\frac m2 \left(1-\frac n2 K\right)
\left[K^2+4\left( 1-\frac {1}{n^2} \right)I^2\right], 
\label{eq:K-3rd}\\
\frac{\partial I}{\partial l} &=& -I\left( nJ+mK\right)- \left(mn-\frac{2m}{n} -\frac {2n}{m} \right) I^2 
\nonumber \\
 & & +\frac{mn}{4} I\left[J^2+K^2 +4\left(1-\frac {1}{n^2} -\frac {1}{m^2} \right) I^2\right].
\label{eq:I-3rd}
\end{eqnarray} 
We note that eq.(\ref{eq:I-3rd}) has a finite $I^2$ term in general.  The coefficient vanishes only if $n=m=2$ with $n$ and $m$ integers.

There are six fixed points of the scaling characterized by eqs.(\ref{eq:J-3rd}) -- (\ref{eq:I-3rd}).  
In correspondence to the SU(2) $\times$ SU(2) model, they are given by
\begin{eqnarray*}
\begin{array}{ll}
{\rm (i)}\ J_c=K_c=I_c=0,  & {\rm (ii)}\ J_c=I_c=0, K_c=2/n, \\ 
{\rm (iii)}\ J_c=2/m, K_c= I_c=0, & {\rm (iv)}\ J_c=2/m, K_c=2/n, I_c=0, \\
{\rm (v)}\ J_c=2/m, K_c=2/n, I_c=I_{mn},  & {\rm (vi)}\ J_c=2/m, K_c=2/n, I_c=1,\\
\end{array}
\end{eqnarray*}
with 
$$ I_{mn}  \equiv -\frac{m^2+n^2}{m^2n^2-m^2-n^2}, $$
the absolute magnitude of which is much smaller than unity for $m,n \gg 1.$
Except for the fixed point (vi), the perturbative renormalization is well controlled because all $J_c,K_c$ and $|I_c|$ are small as compared with unity.  
The case (vi) belongs to the strong-coupling regime and the third-order scaling is unreliable.
In the Coqblin-Schrieffer limit, this fixed point actually flows to the Fermi liquid one in the exact renormalization group. 

Let us study the stability of the fixed points.
The trivial fixed point (i) is unstable as in the case of the standard Kondo model.
The fixed point (ii) constitute a saddle point in the $K-I$ plane; it attracts the renormalization flow parallel to the $K$-axis, but repels it parallel to the $I$-axis.   Similar is the case (iii) with $J$ and $K$ interchanged. 
The linearized scaling equations around the fixed points (iv), (v) and (vi) all contain a part:
\begin{eqnarray}
\frac{\partial J}{\partial l} &=& nm\left (J-\frac 2m\right)
\left[\frac{1}{n^2}+\left( 1-\frac {1}{m^2} \right)I_c^2\right], \\
\frac{\partial K}{\partial l} &=& nm\left(K-\frac 2n\right)
\left[\frac{1}{m^2}+\left( 1-\frac {1}{n^2} \right)I_c^2\right].
\end{eqnarray}
From this set of equations it turns out that the coefficient of the linear term in the right hand side is positive.  
This means that the fixed point attracts the renormalization flow parallel to either $J$- or $K$-axes, independent of the value of $I_c$.  
The flow parallel to the $I$-axis depends on $I_c$ as explained below.

At the fixed point (iv) with $I_c=0$, the remaining linearized scaling equation reads
\begin{equation}
\frac{\partial I}{\partial l} = -\left(\frac mn +\frac nm\right)I.
\end{equation}
The fixed point is unstable since  the coefficient on the right-hand side is positive.

On the other hand, the linearized scaling equation at the fixed point (v) with $I_c =I_{mn}$ is given by
\begin{equation}
\frac{\partial I}{\partial l} = \frac{mn(m^2+n^2)}{m^2n^2-m^2-n^2} (I-I_{mn}).
\end{equation}
Since the coefficient on the right-hand side is positive,  the fixed point is stable.
The fixed point describes a non-Fermi-liquid state.  
Within the third order scaling, this fixed point connects smoothly, 
as the degeneracies $n$ and $m$ decrease,
to the one at $J=K=-I=1$ which was derived in \S\ref{subsec:2 by 2 model}.  
However, the special symmetry in the case of SU(2)$\times$ SU(2) rejects this smooth connection in the exact theory \cite{Kusunose-Kuramoto}.
Thus we suggest that the non-Fermi-liquid fixed point is stable except for the case of SU(2)$\times$ SU(2).
It should be an interesting future problem to see how this strange behavior is related to the symmetry.

\begin{figure}
\begin{center}
\epsfxsize=5cm \epsfbox{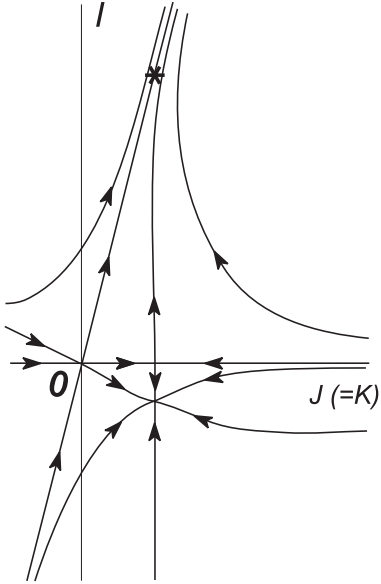}
\end{center}
\caption{
The renormalization flow of the SU($n$)$\times$SU($m$) model in the special case of $n=m$.  The asterisk represents the (fictitious) fixed point given by the third-order scaling.
}
\label{fig:flow}
\end{figure}

Figure \ref{fig:flow} shows qualitatively the flow diagram of the renormalization group in the special case of $n=m \ (\gg 1)$ where the condition $J=K$, if present initially, continues to hold in all steps of renormalization.  
The finite effective interaction in the strong-coupling fixed point at $I_c =1$ has been replaced by the correct flow toward the local Fermi-liquid. 

\section{Summary and Outlook}

In this paper we have proposed a new method to perform renormalization for Kondo-type models.
In contrast to the standard approach which considers the vertex part, our method is concerned with the effective Hamiltonian.  Our approach does not involve infinite-order summation, but still employ the logarithmic approximation.
In this way the multiplicative property in the renormalization is assumed.

In the SU($n$)$\times$SU($m$) model nearly all the fixed points, except a strong-coupling one with $I_c=1$, are shown to be within the reach of the perturbative renormalization.
One of the two non-Fermi-liquid fixed points, which has $I_c =I_{nm}$ and is stable for large $n$ and $m$, becomes unstable against the Fermi liquid one in the special case of the SU(2)$\times$SU(2) symmetry. 
This is due to the unexpected symmetry between $I$ and $-I$.  
Meaning of this hidden symmetry should be identified in a future study. 

The other problem to be pursued further is the renormalization in still higher order.  In particular comparison with exact results obtained by the Bethe Ansatz \cite{Tsvelick,Andrei} and the conformal field theory \cite{Affleck} should be made.  
The latter approach to the multi-channel Kondo model derived the critical value of the exchange interaction as $J_c = 2/(m+2)$, which tends for large $m$ to the value $2/m$ obtained by the third-order renormalization.  
Since the critical value itself is not a universal quantity, the $1/m$ expansion does not necessarily reproduce the value $J_c= 2/(m+2)$.  
However, observable quantities such as the critical exponent must be independent of the method of derivation.
The Rayleigh-Schr\"{o}dinger theory seems to be the simplest approach to compute the higher-order beta function.
The details of the higher-order renormalization will be investigated in a separate paper.

Concerning the physical quantities in the intermediate regime toward the fixed point,  the spin susceptibility can have different temperature dependence from that of the orbital susceptibility.  
Since both spin and orbital fluctuations contribute to the resistivity, conventional argument to relate the transport and magnetic properties via the single Kondo temperature has to be revised in the presence of  an orbital degeneracy.
Recognition of this difference seems to a clue to understanding
the nature of the phase IV of Ce$_x$La$_{1-x}$B$_6$.
As another example of entangled spin and orbital degrees of freedom we mention transition metal oxides, especially manganites.  For these systems we have to modify the model to account for the situation where  the crystal field splitting is much larger than the spin-orbit splitting.
We are now analyzing the SU(2)$\times$SU(2) model quantitatively with use of the numerical renormalization group \cite{Kusunose-Kuramoto}.

\section*{acknowledgments}
This paper is dedicated to Prof. J. Zittartz on the occasion of his sixtieth birthday. 
I would thank Dr. H. Kusunose for useful conversation on the model and the renormalization theory.  


\end{document}